\documentclass{eas}
\usepackage{graphicx}
\newcommand{\dcube}{{${\rm D}^3$}} 
\begin{document}

\title{The Directional Dark Matter Detector (\dcube{})}
\runningtitle{The Directional Dark Matter Detector (\dcube{})}
\author{S.~E.~Vahsen}
\address{University of Hawaii, 2505 Correa Road, Honolulu, HI 96822}
\author{H.~Feng$^1$}
\author{M. Garcia-Sciveres}\address{Lawrence Berkeley National Laboratory, 1 Cyclotron Road, Berkeley, CA 94720}
\author{I.~Jaegle$^1$}
\author{J.~Kadyk$^2$},
\author{Y.~Nguyen$^2$}
\author{M.~Rosen$^1$}
\author{S.~Ross$^1$}
\author{T.~Thorpe$^1$}
\author{J.~Yamaoka$^1$}

\begin{abstract}
Gas-filled Time Projection Chambers (TPCs) with Gas Electron Multipliers (GEMs) and pixels appear suitable for direction-sensitive WIMP dark matter searches. We present the background and motivation for our work on this technology, past and ongoing prototype work, and a development path towards an affordable, 1-$\rm m^3$-scale directional dark matter detector, \dcube. Such a detector may be particularly suitable for low-mass WIMP searches, and perhaps sufficiently sensitive to clearly determine whether the signals seen by DAMA, CoGeNT, and CRESST-II are due to low-mass WIMPs or background.

\end{abstract}
\maketitle

\section{Motivation for Direction Sensitive WIMP Dark Matter Searches}
Dark matter is one of the most compelling mysteries in physics today. Astronomical evidence suggests that there is about five times more cold dark matter than baryonic matter in the universe [\cite{Hinshaw:2008kr}]. A favored scenario is that cold dark matter consists of weakly interacting massive particles (WIMPs). The race is now on to directly detect WIMPs by measuring their collisions with target nuclei in the laboratory. 
In recent years, impressive progress has been made with cryogenic solid- and liquid-phase targets. 
Spin-independent, elastic WIMP/nucleon scattering has been excluded down to cross sections of order $10^{-8}$~pb for WIMP masses from 30 to 100 GeV/$c^2$ [\cite{Censier:2011wd}].
For smaller WIMP-masses, however, the status is much less clear. At least three experiments have observed signals that appear compatible with WIMPs of order 10~GeV/$c^2$. DAMA/LIBRA has claimed positive observation [\cite{Bernabei:2008yi}], the CoGeNT collaboration has reported an ''irreducible excess of bulk-like events'' [\cite{Aalseth:2010vx}], and most recently CRESST-II has observed an excess with significance larger than 4$\sigma$ [\cite{Angloher:2011uu}]. When interpreted as spin-independent, elastic WIMP/nucleon scattering of 10-GeV WIMPs, these results are inconsistent with limits set by XENON100 [\cite{Aprile:2011hi}] and CDMS [\cite{Ahmed:2009zw}]. There are two main possibilities: The excesses seen by DAMA, CoGeNT, and CRESST-II are due to backgrounds that were not fully accounted for (for instance from neutrons [\cite{Ralston:2010bd}]), or they are legitimate dark matter observations, and only appear inconsistent with CDMS and XENON because the dark matter does not scatter off nuclei in the expected manner [\cite{Feng:2011vu}].

In either scenario, it would help to have a detector that can distinguish WIMP-induced and neutron-induced recoils, and that can provide more detailed measurements of each WIMP-recoil candidate event. It was suggested already in 1988 [\cite{Spergel:1987kx}] that unambiguous direct detection of WIMPs may be accomplished with a detector that has directional sensitivity: In the Standard Dark Halo Model an experiment on earth is expected to see a 220 km/s wind of dark matter, due to the rotation of the Milky Way through the dark matter halo. The direction of this wind relative to the detector changes by approximately 90 degrees every twelve hours, due to the tilt of the earth's rotational axis with respect to the galaxy plane.  This 90-degree directional oscillation is the needed smoking gun signal for dark matter detection. It is a much larger effect than the 2\% annual rate oscillation expected due to the motion of the earth around the sun, and no backgrounds (including neutrons) are expected to have this directional signature. 


\section{Existing Efforts on Direction Sensitive WIMP Searches}

\begin{figure}
\begin{center}
\includegraphics[height=40mm]{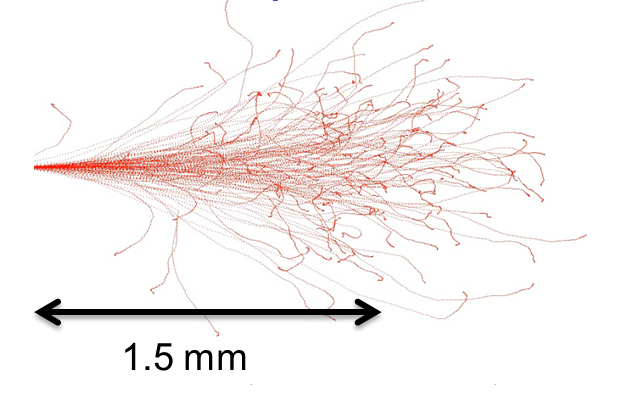}
\label{fig_recoilss} 
\caption{SRIM simulation of 100-keV Fluorine nuclei recoiling in 75 torr of ${\rm CF}_4$ gas. 
A large number of recoils with identical initial position and velocity have been superimposed. For low-mass WIMPs the recoil energy would be lower, so that an even lower gas pressure would be required to obtain reconstructable tracks.}
\end{center}
\end{figure}

The majority of existing WIMP detectors are non-directional, and seek to identify WIMP-induced nuclear recoils by measuring the integrated ionization charge (and/or other signal such as scintillation light and crystal lattice excitations) resulting from the recoils. Directional WIMP detectors also record the three-dimensional ionization trails of the recoils, which are correlated with the direction of the incoming WIMPs. Tracks from WIMP recoils are typically very short, but can be extended to around 1-2~mm in low-pressure gas, as shown in Figure 1. With good detector resolution, such tracks can be measured in detail. 
The DRIFT [\cite{Muna:2007zz}], DMTPC [\cite{Sciolla:2008ak}], MIMAC [\cite{Moulin:2005sx}] and NEWAGE [\cite{Miuchi:2010hn}] collaborations have demonstrated the feasibility of doing so in low-pressure gas-target TPCs. An overview of the different technologies under consideration for directional WIMP detection is given in [\cite{Ahlen:2009ev}].  


\section{The \dcube{} Detection Principle and First Prototype}
\begin{figure}
\begin{center}
\includegraphics[width=80mm]{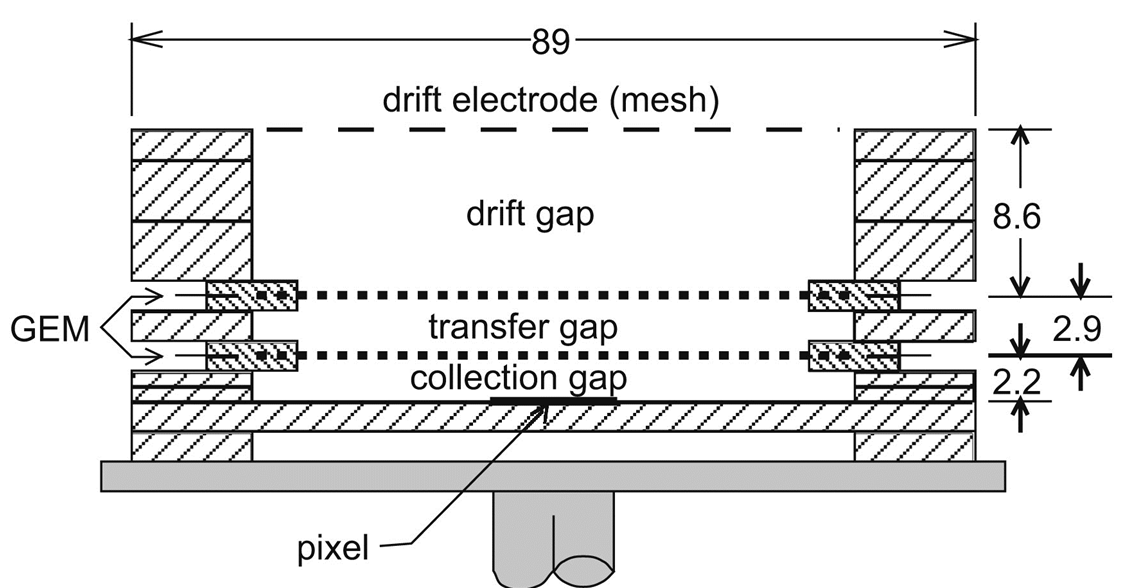}
\end{center}
\caption{First prototype at Lawrence Berkeley National Laboratory [\cite{Kim:2008zzi}]. All dimensions are in mm.}
\label{fig_first_prototype}
\end{figure} 

\begin{figure}
\includegraphics[height=50mm]{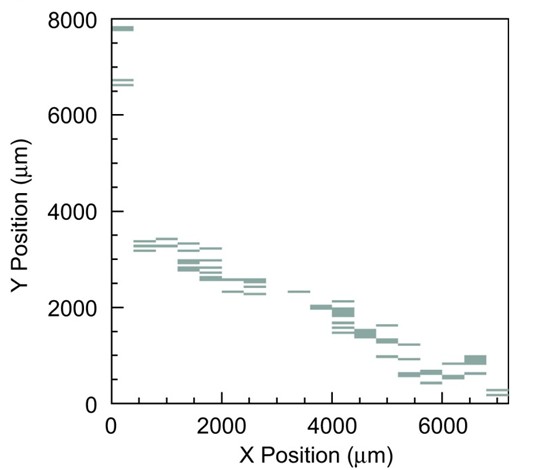}
\includegraphics[height=50mm]{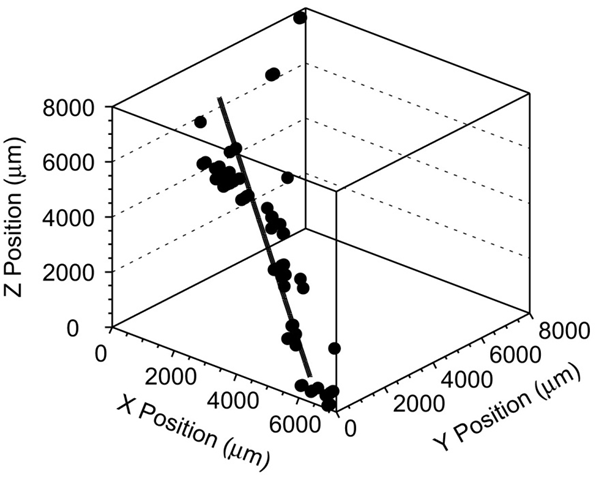}
\caption{A typical cosmic ray event. Left: Each point represents a 50x400 $\mu$m pixel where charge was recorded. Right: The arrival time of charge in each pixel is used to assign a z-position, and the same event is reconstructed in 3D.
\label{fig_cosmic_track}
} 
\end{figure}

 The \dcube{} project is a collaboration between the University of Hawaii and Lawrence Berkeley National Laboratory (LBNL). We are investigating the use of TPCs with Gas Electron Multipliers (GEMs) [\cite{Sauli:1997qp}] and pixel electronics for direction-sensitive WIMP detection and neutron detection. We aim to build a cubic meter size WIMP detector, \dcube{}, in the next years. The detection principle to be used in \dcube{} was first demonstrated in a small LBNL prototype [\cite{Kim:2008zzi}], shown in Figure \ref{fig_first_prototype}, The operating principle is as follows: Primary ionization deposited in the drift gap drifts down to a double layer of GEMs, where it is multiplied by a factor of order $10^3$ to $10^5$. The resulting avalanche charge is collected below the GEMs with an ATLAS FE-I3 pixel chip [\cite{Aad:2008zz}]. The position of pixels where charge is collected gives a two-dimensional trajectory, while the time of charge collection in each pixel provides the third coordinate, as shown in Figure \ref{fig_cosmic_track}. The pixel chip also records the amount of charge collected in each pixel.

Measurements with the LBNL prototype demonstrated  [\cite{Kim:2008zzi}] that TPCs with GEMs and pixels can perform true 3D tracking with ionization measurement (dE/dX), and achieve improved spatial resolution, higher gain, higher efficiency, and lower noise than more typical TPC readout technologies, which employ wire-based readout, optical readouts, or pads. Operating with atmospheric pressure Ar/CO2 gas, we were able to reconstruct cosmic rays with spatial resolution of ~$130~\mu $m (limited by diffusion), corresponding to a readout resolution of ~70 $\mu$m. The noise level in our detector was 120 electrons RMS at room temperature, and we typically operated with pixel thresholds of 1800 electrons and double-GEM gain of 9000. The measured rate of noise hits was initially $<0.2~{\rm minute}^{-1}{\rm cm}^{-2}$, and after masking the three noisiest pixels (out of 2880, for an efficiency of 99.9\%), this rate was reduced to $<0.5~{\rm hour}^{-1}{\rm cm}^{-2}$. Our measurements also suggested that a GEM and pixel readout is capable of detecting all primary ionization - even single electrons - with efficiency near unity.

The performance characteristics of the LBNL prototype are a good match to what is needed in WIMP detection. The spatial resolution is sufficient to reconstruct the short WIMP recoils, while the combination of high gain, low noise, and low threshold promises exceptional sensitivity. In principle, the capability of detecting single, primary electrons with near unit efficiency, should translate into the lowest possible energy threshold when reconstructing nuclear recoils from WIMP/nucleon collisions. More measurements, e.g. of the dependence of pixel noise on GEM gain, are needed to clarify and quantify this. The pixel readout is also likely to result in lower data rates than competing TPC readout approaches. Because every pixel is a self-contained detector with zero suppression, there is negligible data in the absence of signal and physics background, and the detector readout can be highly multiplexed. This may greatly reduce cost and complexity when scaling to a larger detector.

\section{Second Generation Prototypes}

\begin{figure}[h]
\includegraphics[height=55mm]{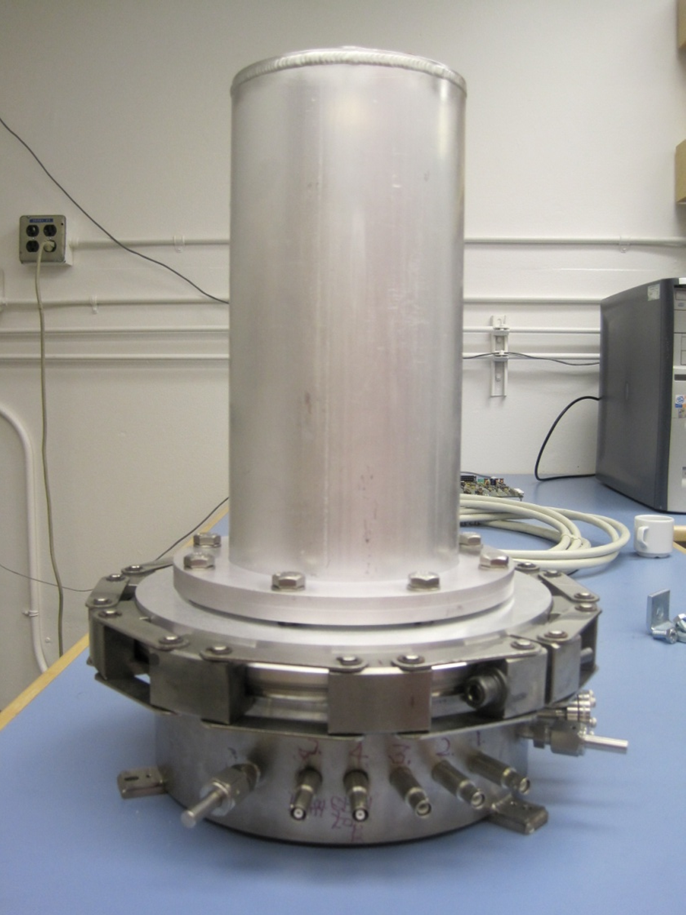}
\includegraphics[height=55mm]{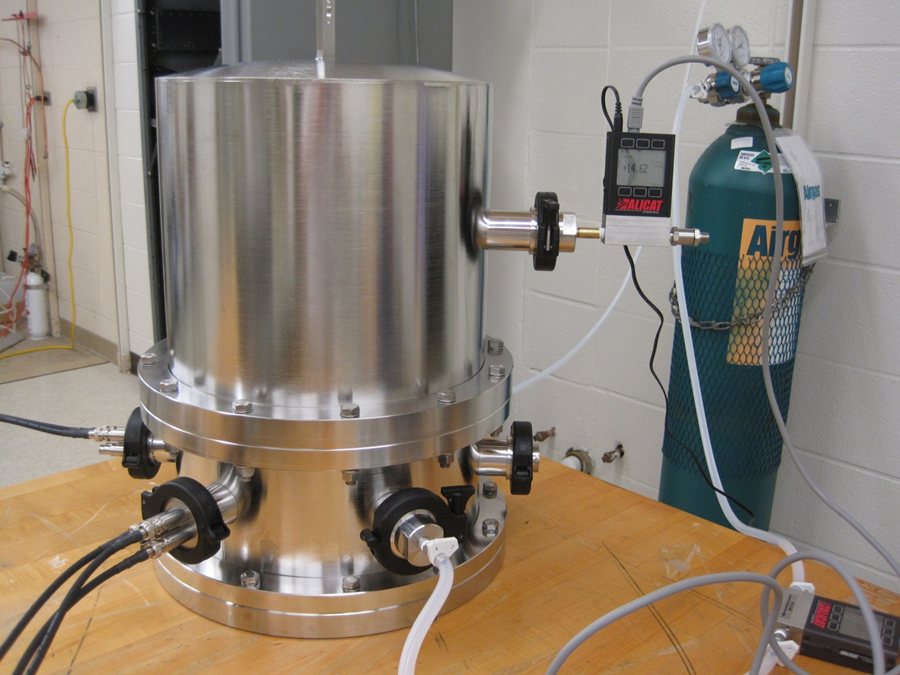}
\caption{Second-generation prototypes at LBNL (left) and the University of Hawaii (right).} 
\label{fig_micro}
\end{figure}

\begin{figure}
\begin{center}
\includegraphics[height=50mm]{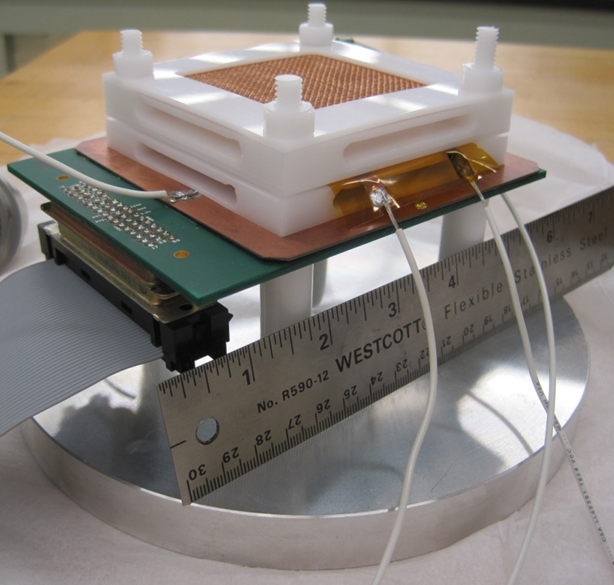}
\includegraphics[height=50mm]{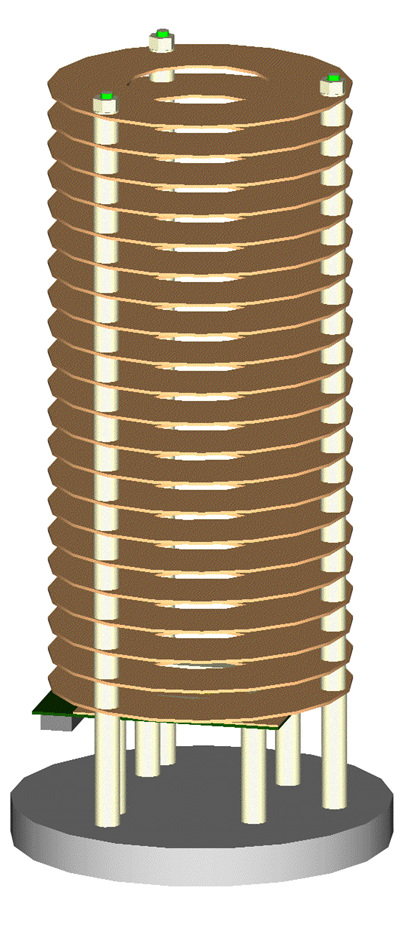}
\end{center}
\caption{Left: Mechanical support structure inside the \dcube{}-micro prototype, with Gas Electron Multipliers (orange, transparent foil) and a Printed Circuit Board (green) with Pixel Chip visible. Right: CAD of the support structure with a field cage installed.}
\label{fig_support}
\end{figure}

In 2010 we embarked on a collaborative effort to build larger prototypes at LBNL and the University of Hawaii, to investigate the feasibility of directional neutron detection and directional WIMP detection in TPCs with GEMs and pixels. The LBNL group extended the vacuum vessel and sensitive volume of the original prototype, and recently re-started data taking. The first Hawaii prototype, \dcube{}-micro, was constructed in 2011 and is currently being commissioned. Figure \ref{fig_micro} shows the two detectors, while Figure \ref{fig_support} shows the current \dcube{}-micro internal support structure, and a CAD model of a copper field cage, that will extend the drift region to 30 cm. Both detectors employ two layers of $(5~{\rm cm})^2$ CERN GEMs, a single ~1-${\rm cm}^2$ ATLAS FE-I3 pixel chip and ATLAS pixel readout electronics. 

\begin{figure}[h]
\begin{center}
\includegraphics[width=80mm]{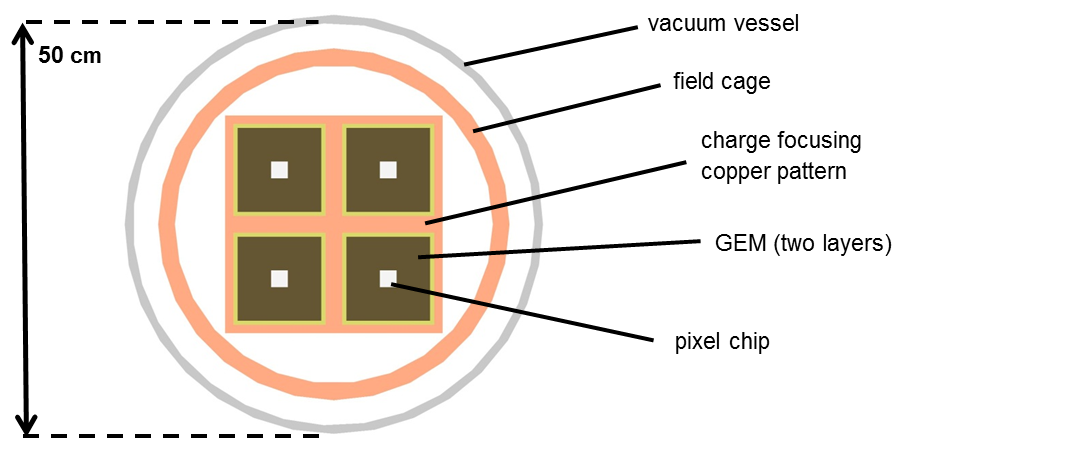}\includegraphics[width=80mm]{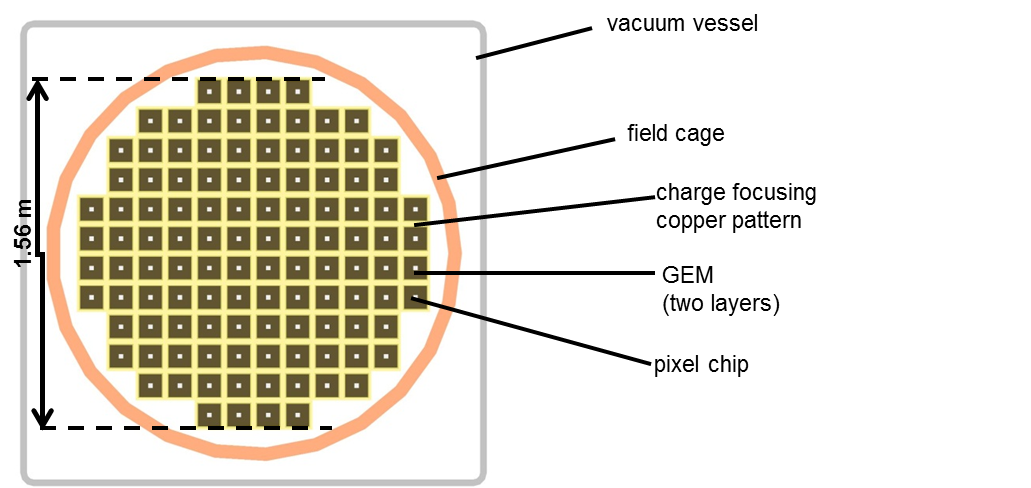}
\end{center}
\caption{Left: Top-view of the planned 12-liter \dcube{}-milli prototype, which implements four charge focusing unit cells of the final \dcube{} inside a common field cage. Right: Concept drawing of the Directional Dark Matter Detector, \dcube{}, top view. Each 30-cm drift layer contains 112 double GEMS, each imaged by a single pixel chip, and has a target volume of 0.484 m$^3$. Multiple layers can be stacked inside the same vaccum vessel, to reach the target volume or budget desired.}
\label{fig_dcube} 
\end{figure}

\section{Feasibility of Large Detectors}

A gas-target WIMP detector will measure of order $1{\rm m}^3$ per kg of target gas, and approximately $(15 \rm m)^3$ for a ton-scale detector. The main challenges in going big are diffusion, cost, and complexity: Realistic detectors designs are limited to a drift length of order 30 cm (for typical pressures of 50 to 100 torr) - beyond that the short tracks from WIMP recoils would be washed out by diffusion. A kg-scale ($\rm m^3$) detector would thus require at least two readout planes of {$1~\rm m^2$, with each plane imaging approximately 30 to 50 cm of drift length. Instrumenting this area fully with pixels would be expensive, at present (using ATLAS FE-I3 or FE-I4 pixels) roughly \$20 per ${\rm cm}^2$, resulting in a total electronics cost of order \$400,000. The construction process would also be very labor intensive, and it is not obvious how to tile the pixel chips, to achieve good area coverage, inside a TPC with high electric field. 

Diffusion may be reduced by transporting the ionization in the TPC with negative ions, which has been demonstrated with GEMs [\cite{Miyamoto:2004dc}] and target gases of interest for directional dark matter detection [\cite{Burgos:2009xm}]. Cost and complexity may be reduced, and the tiling problem eliminated, by focusing the drifting charge with a static electric field, so that a given TPC area is imaged by a smaller pixel area. This takes advantage of the pixels having somewhat better resolution than strictly needed for our application. We have performed initial GARFIELD [\cite{garfield}] simulations that suggest this approach is feasible, but more work is needed. If these results hold up and are confirmed experimentally, then charge focusing may lower the price (and complexity of construction) of a large-scale detector by a factor of 25 to 100, corresponding to linear focusing factors of 5 to 10. In that case a kg-scale (m$^3$) detector could be built with electronics cost between \$4000 and \$16000 - perhaps making even a ton-scale directional detector feasible in the future. Such a detector would be of great interest after WIMPs have been discovered, as it can be used studying the properties of the dark matter halo - "WIMP astronomy".

After detailed measurements with the current, second-generation prototypes, we plan to build a 12-liter prototype, \dcube-milli, which implements four charge focusing unit cells. In each unit cell the charge from a 10x10~cm CERN GEM is focused onto a ~2x2cm ATLAS-FEI4 Pixel Chip, read out with ATLAS USBPix Readout electronics developed by the University of Bonn, Germany. 
We will use \dcube-milli to investigate the limits of charge focusing with and without negative ion drift. 
If \dcube-milli performs as expected, we would then have a validated a detector unit cell that can be repeated multiple times in a larger detector, and proceed to build a 1-$\rm m^3$ detector, \dcube, for deployment underground. A concept drawing is shown in Figure \ref{fig_dcube}.

\section{Expected WIMP Sensitivity}

\begin{figure}
\includegraphics[width=60mm]{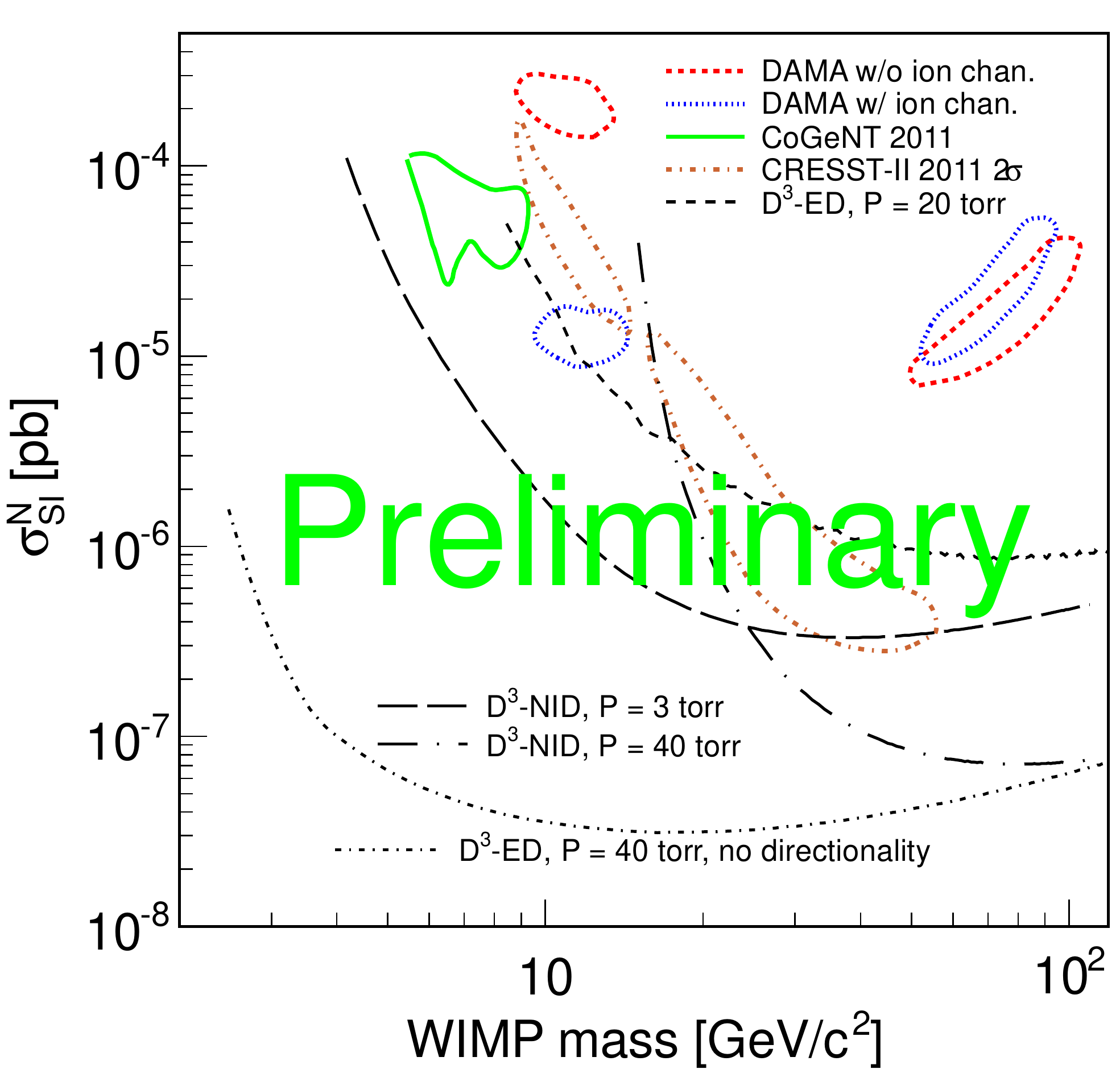}
\includegraphics[width=60mm]{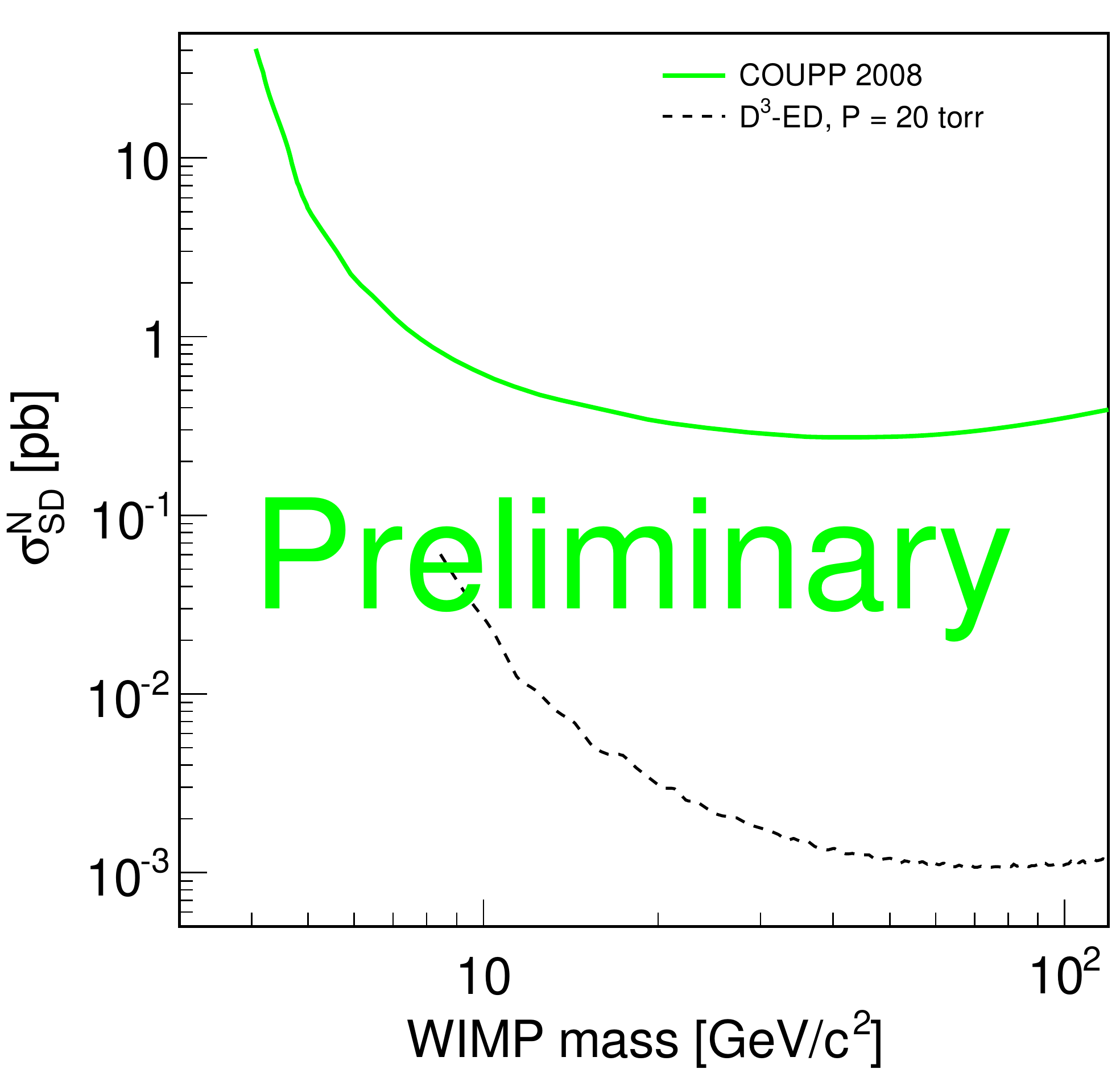}
\caption{Predicted sensitivity of a 3-${\rm m}^3$ $\rm D^3$ to spin-independent (left) and spin-dependent (right) WIMP/nucleon scattering, when operating with low-pressure $\rm{CF}_4$ (labeled ED in plot) or $\rm{CS}_2$ (labeled NID in plot) for three years.} 
\label{fig_sensitivity}
\end{figure}
We have performed preliminary fast simulation studies [\cite{jaegle}] to determine the optimum operating pressure and resulting WIMP sensitivity for a fixed detector geometry; a 3-$\rm m^3$ detector with 33-cm drift length. The results are shown in Figure \ref {fig_sensitivity}. Operating with CF$_4$ gas at 40 torr for three years, we would expect to see several hundred (non-directional) signal events for the WIMP scenarios suggested by CoGeNT and DAMA, and tens even in the lowest-cross-section CRESST-II scenario. The directional sensitivity to such low-mass WIMPs depends strongly on the gas type and pressure. Operating with CF$_4$ gas at twenty torr, we have some directional sensitivity in much of the parameter space, but would really need a larger detector to reach the exposure needed for unambiguous observation of a 12-h oscillation in all scenarios. If we should manage to operate CS$_2$ gas and negative ion drift at 3-4 torr, then already a modest 3-$\rm m^3$ \dcube{} should see of order 10 signal events with directional sensitivity for most of the interesting parameter space, sufficient to observe the 12-h angular oscillation and produce unambiguous evidence for WIMPs.

\section{Conclusion}
Low-pressure gas TPCs with GEM and pixel readout appear suitable for direction-sensitive WIMP searches. The combination of low noise and high single-electron efficiency could make this technology particularly interesting for low-mass WIMP searches. We have presented past and ongoing prototype work, and a development path towards an affordable 1-$\rm m^3$-scale directional dark matter detector, \dcube. Preliminary simulation studies suggest that when operating with negative ion drift, \dcube{} may have sufficient sensitivity to clearly determine whether the signals seen by DAMA, CoGeNT, and CRESST-II are due to low-mass WIMPs or background.




\end{document}